\documentclass[preprint]{JHEP3}
\title{Scattering of Glueballs and the AdS/CFT Correspondence}

\author{Henrique Boschi-Filho\thanks{Work supported in part by CNPq - Brazilian agency.}\\
        Instituto de F\'{\i}sica, Universidade Federal do Rio de Janeiro, Rio de Janeiro, Brazil\\
        E-mail: \email{boschi@if.ufrj.br}}

\author{Nelson R. F. Braga\thanks{Work supported in part by CNPq - Brazilian agency.}\\
       Instituto de F\'{\i}sica, Universidade Federal do Rio de Janeiro, Rio de Janeiro, Brazil\\
       E-mail: \email{braga@if.ufrj.br}}

\abstract{Inspired in the AdS/CFT correspondence one can look for dualities between string theory and non conformal field theories. 
Exact dualities in the non conformal case are intricate but approximations can be helpful in extracting physical results. 
A phenomenological approach consists in introducing a scale corresponding to the maximum value of the axial AdS coordinate. Here we show that this approach can reproduce the scaling of high energy glueball scattering amplitudes and also an approximation for the scalar glueball mass ratios.}

\begin{document}

\section{Strings and strong interactions}

String theory was originally proposed to describe strong interactions.
The motivation was the experimental observation of hadron scattering. 
Considering the scattering of two particles into two particles as in Fig. 1, 
one usually introduces Mandelstam variables:
\begin{equation}s=-(p_1+p_2)^2\quad
  t=-(p_2+p_3)^2\quad
  u=-(p_1+p_3)^2\;,
\end{equation}
satisfying $(s+t+u=\sum m_i^2)$, where $m_i$ are the particle masses.

%%%%%%%%%%%%%%%%%%%%%%%%%%%%%%%%%%%%%%%%%%%%%%%%%%%%%%%%%%
%%%%%%%%%%%%%%%%%%%%%%%%%%%%%%%%%%%%%%%%%%%%%%%%%%%%%%%%%%
%%% figura do espalhamento de duas particulas com momenta 
%%% p_1 e p_2 resultando  em duas particulas com p_3 e p_4
%%%%%%%%%%%%%%%%%%%%%%%%%%%%%%%%%%%%%%%%%%%%%%%%%%%%%%%%%%
%%%%%%%%%%%%%%%%%%%%%%%%%%%%%%%%%%%%%%%%%%%%%%%%%%%%%%%%%%
\
\setlength{\unitlength}{0.04in}
\vskip 1.5cm
\hskip 1cm
{\begin{picture}(0,0)(5,0)
\rm
%%%%%%%%%%%%%%%%%%% Incoming %%%%%%%%%%%%%%%%%%%%%%
\put(-2,15){$p_2$}
\put(3,15){\line(1,-1){5.2}}
\put(-2,0){$p_1$}
\put(3,0){\line(1,1){5.2}}
%%%%%%%%%%%%%%%%%% Interaction %%%%%%%%%%%%%%%%%%%%%
\put(10,7.5){\circle{6}}
%%%%%%%%%%%%%%%%%% Outcomming %%%%%%%%%%%%%%%%%%%
\put(18,15){$p_3$}
\put(17,15){\line(-1,-1){5.2}}
\put(18,0){$p_4$}
\put(17,0){\line(-1,1){5.2}}
%%%%%%%%%%%%%%%%%%%% Equation  %%%%%%%%%%%%%%%%%%%%%%
\hskip 1truecm
\put(25,7){\line(1,0){2}}
\put(25,8){\line(1,0){2}}
%%%%%%%%%%%%%%%%%%%% Canal t %%%%%%%%%%%%%%%%%%%%%%%%
\put(33,15){$p_2$}
\put(38,15){\line(1,-1){5.2}}
\put(33,0){$p_1$}
\put(38,0){\line(1,1){5.2}}
\put(45,6.5){$t$}
\put(43.25,5){\line(0,1){5}}
\put(51,15){$p_3$}
\put(48.5,15){\line(-1,-1){5.2}}
\put(51,0){$p_4$}
\put(48.5,0){\line(-1,1){5.2}}
%%%%%%%%%%%%%%%%%%%%%%   +    %%%%%%%%%%%%%%%%%%%%%%%
\put(58,7){$+$}
%%%%%%%%%%%%%%%%%%%% Canal s %%%%%%%%%%%%%%%%%%%%%%%%
\put(64,15){$p_2$}
\put(67,12.6){\line(1,-1){5.2}}
\put(64,0){$p_1$}
\put(67,2.4){\line(1,1){5.2}}
\put(73.5,9){$s$}
\put(72,7.5){\line(1,0){5}}
\put(83,15){$p_3$}
\put(82,12.6){\line(-1,-1){5.2}}
\put(83,0){$p_4$}
\put(82,2.4){\line(-1,1){5.2}}
%%%%%%%%%%%%%%%%%%% + .... %%%%%%%%%%%%%%%%%%%%%%%%%
\put(90,7){$+$}
\put(97,6.5){$\dots$}
%%%%%%%%%%%%%%%%%%%%%%%%%%%%%%%%%%%%%%%%%%%%%%%%%%%%%
\end{picture}
\vskip .5 cm
\centerline{Fig. 1:
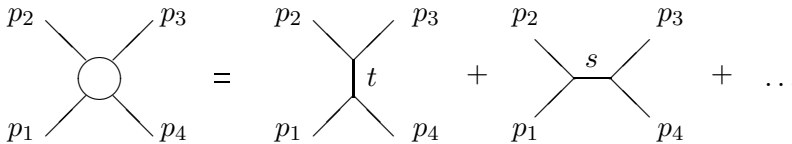
{\it Two-two scattering in terms of Mandelstam variables.}}
\vskip .5cm
%%%%%%%%%%%%%%%%%%%%%%%%%%%%%%%%%%%%%%%%%%%%%%%%%%%%%%
%%%%%%%%%%%%%%%%%%%%%%%%%%%%%%%%%%%%%%%%%%%%%%%%%%%%%%

Taking into account the observed properties known at that time, 
Veneziano proposed an ingenious formula for the hadron scattering amplitude  
\cite{Green:1987sp}
\begin{equation}
{\cal A}(s,t)=\frac{\Gamma(-\alpha(s))\;\Gamma(-\alpha(t))}
{\Gamma(-\alpha(s)-\alpha(t))}\;,
\end{equation}
where $\;\;\alpha(s)=\alpha(0)+{\alpha\,}^\prime s\;\;$ are the Regge's trajectories. 
This amplitude can be deduced from string theory which also predicts 
a relation between mass and angular momentum 
\begin{equation}
({m})^2=\frac{J-\alpha(0)}{{\alpha\,}^\prime}\;,
\end{equation}
in agreement with experimental data for hadrons with spin $J$.

The Veneziano amplitude reproduces the high energy behavior of hadronic amplitudes in the so called Regge regime corresponding to $\; s\to \infty\;$ with $\; t\;$ fixed (and vice-versa). However, considering hadronic high energy scattering with fixed angles corresponding to $\; s,\; t\;\to \infty\;$ with the ratio $\; s/t\;$ fixed,  the Veneziano's amplitude presents a soft behavior 
\begin{equation}
{\cal A}_{Veneziano}\;\quad\sim\quad e^{-s}\;,
\end{equation}
while experiments show a hard behavior 
\begin{equation}
{\cal A}_{Experimental}\;\quad\sim\quad {s}^{-constant}\,.
\end{equation}

\noindent
This experimental scaling is reproduced from QCD as shown by   
Brodsky and Farrar \cite{Brodsky:1973kr}, 
Matveev {\sl et. al.} \cite{Matveev:1973ra}.

Another puzzle for string theory, regarding strong interactions, is the difficulty in reproducing the results of deep inelastic lepton-hadron scattering (Fig. 2) as for instance the Bjorken scaling. These experiments showed also that at high energy protons and neutrons are made of point-like objects later identified with quarks. 

%%%%%%%%%%%%%%%%%%%%%%%%%%%%%%%%%%%%%%%%%%%%%%%%%%%%%%%%%%
%%%%%%%%%%%%%%%%%%%%%%%%%%%%%%%%%%%%%%%%%%%%%%%%%%%%%%%%%%
%%% figura do espalhamento profundamente inelastico   %%%%
%%%%%%%%%%%%%%%%%%%%%%%%%%%%%%%%%%%%%%%%%%%%%%%%%%%%%%%%%%
%%%%%%%%%%%%%%%%%%%%%%%%%%%%%%%%%%%%%%%%%%%%%%%%%%%%%%%%%%
\
\setlength{\unitlength}{0.06in}
\vskip 2.5 cm
\hskip 2.5cm {\begin{picture}(0,0)(-15,0)
\rm
%%%%%%%%%%%%%%%%%%% Lepton %%%%%%%%%%%%%%%%%%%%%%
\put(0,15){$e^{-}$}
\put(3,15){\line(2,-1){7}}
\put(18,15){$e^{-}$}
\put(17,15){\line(-2,-1){7}}
%%%%%%%%%%%%%%%%%%%  Foton  %%%%%%%%%%%%%%%%%%%%%%%
\put(9.5,8){$\gamma$}
\bezier{300}(10,11.5)(10.2,10.7)(11,10.5)
\bezier{300}(11,10.5)(11.8,10.3)(12,9.5)
\bezier{300}(12,9.5)(12.2,8.7)(13,8.5)
\bezier{300}(13,8.5)(13.8,8.3)(14,7.5)
%%%%%%%%%%%%%%%%%%   Proton  %%%%%%%%%%%%%%%%%%%%%%%
\put(-3,3){ Proton}
\put(3,0){\line(2,1){10.5}}
%%%%%%%%%%%%%%%%%% Interaction %%%%%%%%%%%%%%%%%%%%%
\put(16,6){\circle{5}}
%%%%%%%%%%%%%%%%%%   Hadrons %%%%%%%%%%%%%%%%%%%%%%%
\put(27,-2){ Hadrons}
\put(18.5,5.5){\line(3,-1){8}}
\put(18.3,5){\line(2,-1){8}}
\put(18,4.5){\line(3,-2){7.5}}
\put(17.5,3.8){\line(1,-1){6}}
%%%%%%%%%%%%%%%%%%%%%%%%%%%%%%%%%%%%%%%%%%%%%%%%%%%%%
\end{picture}
\vskip .5 cm
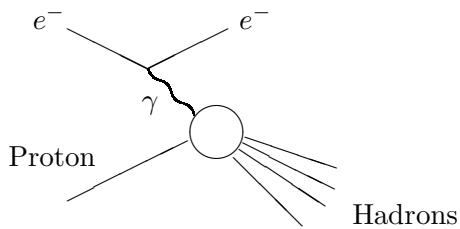
\centerline{Fig. 2: {\it Lepton-Hadron scattering.}}
\vskip .5 cm
%%%%%%%%%%%%%%%%%%%%%%%%%%%%%%%%%%%%%%%%%%%%%%%%%%%%%%
%%%%%%%%%%%%%%%%%%%%%%%%%%%%%%%%%%%%%%%%%%%%%%%%%%%%%%

Then, string theory {\sl in flat spacetime} does not describe correctly the strong interactions of hadrons.

%%%%%%%%%%%%%%%%%%%%%%%%%%%%%%%%%%%%%%%%%%%%%%%%%%%%%%%%%%%%%%%%%%%%%
%%%%%%%%%%%%%%%%%%%%%%%%%%%%%%%%%%%%%%%%%%%%%%%%%%%%%%%%%%%%%%%%%%%%%
%%%%%%%%%%%%%%%%%%%%%%%%%%%%%%%%%%%%%%%%%%%%%%%%%%%%%%%%%%%%%%%%%%%%%

\section{AdS/CFT correspondence and holographic map}

In 1974 't Hooft studied in a remarkable work \cite{'tHooft:1973jz} 
the perturbative expansion of U(N) Gauge theories in the limit $N \to\infty$, 
with $g^2N$ fixed. He showed that planar diagrams with quarks at the edges 
dominate the perturbative series with parameter $1/N$. 
The topological structure of the $1/N$ series is identical to that of the dual model (strings) with $1/N$ as the coupling constant.

Recently, Maldacena \cite{Maldacena:1997re,Aharony:1999ti} proposed that 
compactifications of M/string theory on various Anti-de Sitter spacetimes 
are dual to various  conformal  field theories.
The large $N$ limit for ${\cal N} =4$ super-Yang-Mills theory in four dimensions 
is equivalent to  type IIB superstrings in five dimensional Anti-de Sitter space times a 
5 dimensional sphere (AdS$_5\times$S$^5$). 
This is known as the AdS/CFT correspondence. 

It was shown recently by Polchinski and Strassler that the hard scattering behavior of strong interactions can be obtained from string theory based on the AdS/CFT correspondence\cite{Polchinski:2001tt}. They have also studied deep inelastic scattering using this framework \cite{Polchinski:2002jw}.

In order to obtain a map between AdS bulk and boundary let us consider 
type IIB string theory approximated at low energy $(<< 1/\sqrt{\alpha^\prime})\,$  
by supergravity action: 
\begin{equation}
S \,=\, {1\over 2 \kappa^2} \int d^{10}x \sqrt{G}  \Big[
{\cal R} + G^{MN} \partial_M \Phi \partial_N \Phi  \,+\, ....\Big]\;,
\end{equation}
 
\noindent 
where $G^{MN}$ is the 10-d metric, ${\cal R}$ is the scalar curvature,  
$\Phi$ is the  dilaton and $\kappa \sim g (\alpha^\prime)^2 $. 
The 10-d spacetime is identified with AdS$_{5}\,\times S^5$
\begin{equation}
\label{metric}
ds^2=\frac {R^2 }{ z^2}\Big( dz^2 \,+(d\vec x)^2\,
- dt^2 \Big) \,+ \,R^2 d\Omega_5^2 \,\,,
 \end{equation}
 
\noindent 
where $R$ is the AdS radius and $\Omega_5$ describe the $S^5$ sphere.  

To set an energy scale we consider just a slice of the AdS space:
\begin{equation}\;0\;\le\; z\; \le\; z_{max}\;.\end{equation}
 
Assuming the dilaton to be in s-wave state (so that $S^5$ coordinates are irrelevant), 
the supergravity action in the slice is:
\begin{equation}
\label{SGA}
S ={\pi^3 R^8 \over 4 \kappa^2} \int d^4x \int_0^{z_{max}} {dz\over z^3} 
\Big[{\cal R} + ( \partial_z \Phi )^2 + \eta^{\mu\nu} \partial_\mu \Phi  
\partial_\nu \Phi+ ....\Big]\;,
\end{equation}
 
\noindent where $\eta^{\mu\nu}$ is the Minkowiski 4-d metric. 
The corresponding solution assuming the field to vanish at $z = z_{max}$ is 
\cite{Boschi-Filho:2000vd,Boschi-Filho:2000nh}
\begin{equation}
\Phi(z,\vec x,t) = \sum_{p=1}^\infty \,
\int { d^3 k \over (2\pi)^{3}}\,
{z^{2} \,J_2 (u_p z ) \over z_{max} w_p(\vec k ) 
\,J_{3} (u_p z_{max} ) }
\lbrace { { a}_p(\vec k )\ }
 e^{-iw_p(\vec k ) t +i\vec k \cdot \vec x}\,
\,+\,\,h.c.\rbrace\,,
\end{equation}

\noindent where  $\quad w_p(\vec k ) \,=\,\sqrt{ u_p^2\,+\,{\vec k}^2}\;$  
 and 
\begin{equation}
u_p z_{max}\,=\, \chi_{_{2\,,\,p}}\;,
\end{equation}

\noindent such that  $ J_2 (\chi_{_{2\,,\,p}} )=0$. 
The creation and annihilation operators satisfy the algebra
\begin{equation}
\Big[ { a}_p(\vec k )\,,\,{ a}^\dagger_{p^\prime}({\vec k}^\prime  )
\Big]\,=\, 2\, (2\pi)^{3} w_p(\vec k )   
\delta_{p\,  p^\prime}\,\delta^{3} (\vec k -
{\vec k}^\prime )\,.
\end{equation}
 
\noindent On the AdS boundary $ z = 0$ we consider massive composite operators $\Theta_i(\vec x,t)\,$ associated with glueballs. The corresponding 
creation-annihilation operators for asymptotic states are assumed to satisfy 
the algebra
\begin{equation}
\Big[ { b}_i( \vec K )\,,\,{ b}_j^\dagger ({ \vec K }^\prime  )
\Big]\,=\, 2 \delta_{ij} \, (2\pi)^{3} \, w_i( \vec K ) \,\delta^3 ( \vec K -
{ \vec K}^\prime ) \,,
\end{equation}

\noindent where $\quad w_i(\vec K ) = \sqrt{ {\vec K}^2 + \mu_i^2}\;$ and $\mu_i$ is the mass of the field $\Theta_i$.

Now let us work out a one to one map between AdS bulk field $\Phi$  
and boundary fields $\Theta_i$, in particular relating their creation annihilation operators \cite{Boschi-Filho:2001km}. Note that the dilaton $\Phi$ lives in a 5-d space while $\Theta$ lives in a 4-d space, but the dilaton has a discrete spectrum associated with the $u_p$ modes.

Let us impose that creation-annihilation operators of both theories be related by 
\begin{eqnarray}
\label{ab}
k\,{ a}_i ( \vec  k ) 
&=& K \,{ b}_i( \vec K  )\;, \nonumber\\
k\,{ a}^\dagger_i ( \vec k ) 
&=& K\,{ b}_i^\dagger ( \vec K  )\,.
\end{eqnarray}

\noindent To preserve the commutation relations of both theories 
we find an equation which solution reads  
\begin{equation}
\label{completo}
k = {u_i \over 2} 
\,\left[ \,{ {\cal E}  +\sqrt{{\cal E}^2 + \mu_i^2 } 
			\over  K + \sqrt{K^2 + \mu_i^2}}
- { K + \sqrt{K^2 + \mu_i^2}\over {\cal E} 
+\sqrt{{\cal E}^2 + \mu_i^2 }}\,\right]\,,
\end{equation}

\noindent where $\cal E$ is an UV cuttoff for the boundary theory.

\section{Glueball scattering amplitudes}

The AdS/CFT prescriptions relate the bulk dilaton field to boundary scalar 
glueball operators. Let us see how the bulk-boundary map of the previous section can be used to reproduce the high energy scaling of scalar glueball amplitudes.

First we relate the mass $\mu_1\equiv\mu$ of the lightest glueball with the size of the AdS slice:
\begin{equation}
z_{max} \,\sim\, {1\over \mu}\;,
\end{equation}
 
\noindent we see that $z_{max}$ corresponds to an IR cutoff of the boundary theory. Since  $u_1\,z_{max} \, \sim \, 1$ this implies that  $u_1 \sim \mu $.
Also, as ${\cal E}_1\,$ is an UV cutoff then the high energy scattering momenta 
$\,  K \,$ of the glueballs satisfy  $\mu \ll K \ll {\cal E}$. 
This way, the relation between bulk $k$ and boundary $K$ momenta can be approximated by
\begin{equation}
\label{Kk}
k \,\,\approx\,\, { {\cal E}\, \mu  \over 2 \, K}\,.
\end{equation}

\noindent
Note that this relation combined with the condition    
$\mu \ll K \ll {\cal E} $ implies 
$$\mu \ll k \ll {\cal E}\;.$$
  
\noindent
Setting the string energy scale to the UV boundary cutoff, 
$\sqrt{\alpha^\prime}\,\sim 1/ {\cal E}\;,$ we find that bulk momenta $k$ satisfy 
$\mu \ll k \ll 1/\sqrt{\alpha^\prime}\;$ 
which is consistent with the fact that we are taking the low energy (supergravity) limit of string theory.

Let us consider the scattering of two particles into $m$ 
particles all in the bulk with axial momentum $u_1$ (see Fig. 3).

%%%%%%%%%%%%%%%%%%%%%%%%%%%%%%%%%%%%%%%%%%%%%%%%%%%%%%%%%%
%%%%%%%%%%%%%%%%%%%%%%%%%%%%%%%%%%%%%%%%%%%%%%%%%%%%%%%%%%
%%% figura do espalhamento de duas particulas entrando  %%
%%%%%%%%%%%%%%% resultando  em m particulas saindo  %%%%%%
%%%%%%%%%%%%%%%%%%%%%%%%%%%%%%%%%%%%%%%%%%%%%%%%%%%%%%%%%%
%%%%%%%%%%%%%%%%%%%%%%%%%%%%%%%%%%%%%%%%%%%%%%%%%%%%%%%%%%
\
\setlength{\unitlength}{0.05in}
\vskip 2cm
\hskip 6cm
{\begin{picture}(0,0)(5,0)
\rm
%%%%%%%%%%%%%%%%%%% Incomming %%%%%%%%%%%%%%%%%%%%%%
\put(-4,14){$\vec k_2,u_1$}
\put(3,15){\line(1,-1){5.2}}
\put(3,15){\vector(1,-1){3.2}}
\put(-4,-1){$\vec k_1,u_1$}
\put(3,0){\line(1,1){5.2}}
\put(3,0){\vector(1,1){3.2}}
%%%%%%%%%%%%%%%%%% Interaction %%%%%%%%%%%%%%%%%%%%%
\put(10,7.5){\circle{6}}
%%%%%%%%%%%%%%%%%% Outcomming %%%%%%%%%%%%%%%%%%%
\put(18,15){$\vec k_3,u_1$}
\put(17,15){\line(-1,-1){5.2}}
\put(17,15){\vector(1,1){.2}}
\put(18,-2){$\vec k_{m+2},u_1$}
\put(17,0){\line(-1,1){5.2}}
\put(17,0){\vector(1,-1){.2}}
\put(18,7.5){\line(-1,0){5.2}}
\put(18,7.5){\vector(1,0){.2}}
\put(18,11){\line(-2,-1){5.2}}
\put(18,11){\vector(2,1){.2}}
\put(18,4){\line(-2,1){5.2}}
\put(18,4){\vector(2,-1){.2}}
%%%%%%%%%%%%%%%%%%%%%%%%%%%%%%%%%%%%%%%%%%%%%%%%%%%%%
\end{picture}
\vskip .5 cm
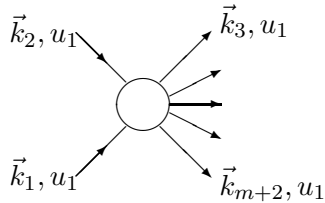
\centerline{Fig. 3: {\it The scattering of 2 particles into $m$ particles.}}
\vskip .5cm
%%%%%%%%%%%%%%%%%%%%%%%%%%%%%%%%%%%%%%%%%%%%%%%%%%%%%%
%%%%%%%%%%%%%%%%%%%%%%%%%%%%%%%%%%%%%%%%%%%%%%%%%%%%%%

 The $S$ matrix is  
\begin{eqnarray} 
S_{Bulk} &=&  \langle \, {\vec k}_3\,,u_1;\,...; \,{\vec k}_{m+2}\,,u_1\,
;\,out \vert {\vec k}_1,\,u_1\, 
;\,{\vec k}_2 \,,u_1;in\,\rangle \nonumber\\
&=&  \langle \, 0\,\vert \,{ a}_{out} ( {\vec k}_3 )\,... \,{ a}_{out}
 ({\vec k}_{m+2}) \,  { a}^+_{in }
( {\vec k}_1) \,{ a}^+_{in} ({\vec k}_2) \,\vert \, 0\, \rangle \;,
\end{eqnarray}

\noindent where ${ a} \equiv { a}_1$ and the $in$ and  $out$ states are defined as 
$ \vert \vec k \,,\,u_1\,\rangle \,=\, { a}^+ (\vec k ) \vert 0 \rangle \,$.

Using the map between creation-annihilation operators of bulk and boundary theories and considering fixed angle scattering, 
$ k_i \,=\, \gamma_i\; k\quad $ and  $\quad K_i \,=\, \Gamma_i\; K$, 
where $\gamma_i $ and  $\Gamma_i$ are constants ($i\,=\,1,2,...,m+2)\,$ 
we have 
\begin{eqnarray}
S_{Bulk} &\sim&   \langle  0 \vert \,{ b}_{out} ( {\vec K}_3 )\,... \,
\,{ b}_{out} ({\vec K}_{m+2})   
\,{ b}^+_{in }( {\vec K}_1)\, { b}^+_{in} ({\vec K}_2) \vert 0 \rangle  
\,\Big({ K \over k}\Big)^{m + 2 } \nonumber\\
&&\sim\, \langle  \,{\vec K}_3 \,,... \,{\vec K}_{m+2},\,out \,\vert \,{\vec K}_1 \,,
{\vec K}_2 \,,in\,\rangle \, \Big( { K \over k} \Big)^{m+ 2} K^{(m+2)(d-1) }\,,
\end{eqnarray}

\noindent   where $d$ is the dimension of the boundary composite operators and ${ b} \equiv { b}_1$ .  The corresponding $in$ and $out$ states are 
$ \vert \vec K \,\rangle \,\cong\, K^{  1 - d } { b}^+ (\vec K ) \vert 0 \rangle \,,$
since $K \gg \mu$.

Using now the relation (\ref{Kk}) between bulk and boundary momenta we find 
\begin{equation}
S_{Bulk} \, \sim \,  
 S_{Bound.} \,\,\Big( {\sqrt{\alpha^\prime} \over \mu }\Big)^{m+2} \,\, K^{(m+2)(1+d)}\;.
\end{equation}

The scattering amplitudes ${\cal A}$ are related to the $S$ matrix by 
\begin{eqnarray}
S_{Bulk} &=& {\cal A}_{Bulk} \, 
\delta^4 (  k_1^\rho +  k_2^\rho -  k_3^\rho - \,...\,- 
 k_{m+2}^\rho )\;,
\nonumber\\
 S_{Bound.} &=& {\cal A}_{Bound.} \, \delta^4 ( K_1^\rho +  K_2^\rho 
- K_3^\rho - ... -K_{m+2}^\rho  )\;, 
\end{eqnarray}

\noindent  so we find 
 \begin{eqnarray}
\label{Mb}
{\cal A}_{Bound.} &\sim& {\cal A}_{Bulk}\,\,  S_{Bound.}\,\, (\, S_{Bulk}\,)^{-1} 
\Big( { K\over k} \Big)^4 \nonumber\\
&\sim&  {\cal A}_{Bulk}\, \,K^{8 -  (m+2)(d + 1)  } \,\,\,
\Big( {\sqrt{\alpha^\prime} \over \mu }\Big)^{2 - m}\,\,.
\end{eqnarray}

The bulk scattering amplitude can be estimated from supergravity action in the AdS slice (\ref{SGA}) using dimensional arguments. Note that  $\, R^8 / \kappa^2\, $ is dimensionless. The dimensionfull parameters are $z_{max}\,\,\sim\,1/\mu\,$ and the Ricci scalar $ {\cal R} \,\sim \, 1/R^2\,$. But  $\mu\,\ll\,k$, so the relevant contribution to the amplitude at high energy will not depend on $z_{max}$. 
Further, choosing  the AdS radius $R$ such that $ 1/R \ll k $ (small curvature) 
we can disregard the contribution from the Ricci scalar to the amplitude. 
So, the only relevant bulk dimensionfull parameter is the momentum $k$.

Taking into account the normalization of states  $\vert k, u_1\rangle\,$ one finds that 
$\,{\cal A}_{Bulk}\,$ has dimension  of [Energy]$^{4 - n}$, where  $n=2+m$ is the total number of scattered particles. Then, 
\begin{equation}
{\cal A}_{Bulk}\,\sim\, k^{ 2 - m }\,\,.
\end{equation}

Using again the relation (\ref{Kk}) between bulk and boundary momenta and substituting the above equation in (\ref{Mb}), we find 
\begin{equation}
{\cal A}_{Bound.} \,\sim \,K^{4 - \Delta } \,,
\end{equation}

\noindent where  $\Delta = ( m + 2) d $ is the total scaling dimension of the 
scattered particles associated with glueballs on the 4-d boundary. 

Considering that $ K \sim \sqrt{ s} $ we find \cite{Boschi-Filho:2002zs}
\begin{equation}
 {\cal A}_{Bound.} \,\sim \,s^{2  - \Delta/2 } \,, 
\end{equation}
which is the QCD hard scattering result \cite{Brodsky:1973kr,Matveev:1973ra}
that was reproduced from AdS / CFT correspondence in \cite{Polchinski:2001tt}. 
Other interesting related discussions can be found in 
\cite{Brower:2002er,Andreev:2002aw,
Brodsky:2003px,deTeramond:2004qd,Andreev:2004sy}.

This phenomenological approach of introducing a mass scale by taking an AdS slice 
can also de used to calculate glueball mass ratios 
\cite{Boschi-Filho:2002vd,Boschi-Filho:2002ta}. 
This way we get a relation between the dilaton axial modes $u_p$ and the glueball masses $\mu_p$:
\begin{equation}
{u_p \over \mu_p }\,=\,constant\,\,.
\end{equation}
 
\noindent So the glueball masses can be related to the zeros of the Bessel
functions by
\begin{equation}
\label{QCD4}
{ \mu_p\over \mu_1 }\,=\,{\chi_{2\,,\,p}\over \chi_{2\,,\,1}}\,\,.
\end{equation}
These results are in good agreement with lattice and supergravity calculations as shown in \cite{Boschi-Filho:2002vd,Boschi-Filho:2002ta}.

%%%%%%%%%%%%%%%%%%%%%%%%%%%%%%%%%%%%%%%%%%%%%%%%%%%%%%%%%%%%%%%%%%%%%%%
%%%%%%%%%%%%%%%%%%%%%%%%%%%%%%%%%%%%%%%%%%%%%%%%%%%%%%%%%%%%%%%%%%%%%%%
%%%%%%%%%%%%%%%%%%%%%%%%%%%%%%%%%%%%%%%%%%%%%%%%%%%%%%%%%%%%%%%%%%%%%%%
%%%%%%%%%%%%%%%%%%%%%%%%%%%%%%%%%%%%%%%%%%%%%%%%%%%%%%%%%%%%%%%%%%%%%%%

\end{document}